\documentclass[twocolumn,superscriptaddress,amsfont,amssymb,amsmath, showpacs,balancelastpage, nofootinbib]{revtex4-1}
\usepackage{graphicx,longtable,color,array}
\usepackage[hidelinks]{hyperref}
\usepackage[usenames,dvipsnames]{xcolor}
\usepackage{amssymb,mathrsfs}
\usepackage{mathtools}
\usepackage{bm}
\usepackage{amsmath}
\usepackage{tikz}
\usepackage[caption=false,justification=justified]{subfig}
\usepackage[compat=1.1.0]{tikz-feynman}
\usepackage{booktabs,longtable,ctable,multirow}
\usepackage{exscale,relsize}
\usepackage[normalem]{ulem}
\usepackage{enumerate}
\usepackage{colortbl} 
\usepackage[dvipsnames]{xcolor} 
\usepackage{pifont} 
\usepackage{adjustbox}

\allowdisplaybreaks

\newcommand{\be}{\begin{equation}}      
\newcommand{\ee}{\end{equation}}

\newcommand{\mpl}{m_{\text{Pl}}}

\newcommand{\vv}{{\bf v}}
\newcommand{\vb}{{\bf b}}
\newcommand{\vk}{{\bf k}}
\newcommand{\vq}{{\bf q}}

\newcommand{\vn}{{\bf n}}
\newcommand{\ve}{{\bf e}}

\def\dd{\delta\!\!\!{}^-\!}
\newcommand{\uu}{{\cal U}}

\begin{document}

\title{Gravitational Bremsstrahlung\\ in the Post-Minkowskian Effective Field Theory}
\author{Stavros Mougiakakos}  
\author{Massimiliano Maria Riva} 
\author{Filippo Vernizzi}
\affiliation{Institut de physique th\' eorique, Universit\'e  Paris Saclay CEA, CNRS, 91191 Gif-sur-Yvette, France}
        \date{\today}

\begin{abstract}

We study the gravitational radiation emitted during the scattering of two spinless bodies in the post-Minkowskian Effective Field Theory approach. We  derive the 
conserved stress-energy tensor linearly coupled to gravity and  the classical probability amplitude of graviton emission at leading and next-to-leading order in the Newton's constant $G$.
The amplitude can be expressed in compact form as one-dimensional integrals over a Feynman parameter involving Bessel functions.
We use it to  recover the leading-order radiated angular momentum  expression. 
Upon expanding it  in the relative velocity between the two bodies $v$, we compute the  total four-momentum radiated into gravitational waves at leading-order in $G$ and up to an order $v^8$, finding agreement with what was recently computed using scattering amplitude methods. 
Our results also allow us to investigate the zero frequency limit of the emitted energy spectrum.

\end{abstract}

\maketitle

\section{Introduction}

The understanding of the  dynamics of binary systems and their gravitational wave emission has been crucial for the extraordinary discovery of LIGO/Virgo \cite{Abbott:2016blz,TheLIGOScientific:2017qsa}. This field has recently received a renewed attention, particularly in the application of the so-called post-Minkowskian (PM) framework 
\cite{Bertotti:1956pxu,Bertotti:1960wuq,Havas:1962zz,Westpfahl:1979gu,Portilla:1980uz,Bel:1981be,Westpfahl:1985tsl,Damour:2016gwp,PhysRevD.97.044038,Blanchet:2018yvb}, which consists of
expanding the gravitational dynamics in the  Newton's constant $G$ while keeping the velocities fully relativistic. This is complementary to the post-Newtonian approach (see \cite{Blanchet:2013haa,Blanchet:2018hut} and references therein), where  one expands  in both velocity and $G$, since in a bound state these two are related by  the virial theorem. 

Recently, 
progress has been  made within the PM framework thanks to the application of several complementary approaches:  in particular
the effective one-body method \cite{Damour:2016gwp,PhysRevD.97.044038,Antonelli:2019ytb,Damour:2019lcq}, the use of scattering amplitude technics, such as the double copy \cite{Bern:2008qj,Bern:2010ue,Bern:2019prr}, generalized unitarity \cite{Bern:1994zx,Bern:1994cg,Britto:2004nc} and effective field theory (EFT)  \cite{Neill:2013wsa,Bjerrum-Bohr:2013bxa,Luna:2017dtq,Bjerrum-Bohr:2018xdl,Kosower:2018adc,Cheung:2018wkq,Cristofoli:2019neg,Cristofoli:2020uzm} (see \cite{Veltman:1975vx,DeWitt:1967yk,DeWitt:1967ub,DeWitt:1967uc,PhysRevD.50.3874,PhysRevD.67.084033,Iwasaki:1971iy,Iwasaki:1971vb,Mougiakakos:2020laz} for the quantum field theoretic description of gravity), and  worldline EFT approaches  \cite{Goldberger:2016iau,Goldberger:2017vcg,Kalin:2020mvi,Loebbert:2020aos,Mogull:2020sak}. 
These developments concern the scattering of  unbound states but results can be extended to bound states by applying an  analytic continuation between hyperbolic and elliptic motion \cite{Kalin:2019rwq,Kalin:2019inp}.
The progress has  addressed the conservative  binary dynamics up to 3PM order \cite{Bern:2019nnu,Bern:2019crd,Cheung:2020gyp,Kalin:2020fhe}, as well as  tidal   \cite{Kalin:2020lmz,Bern:2020uwk,Cheung:2020sdj,AccettulliHuber:2020oou,Haddad:2020que,Aoude:2020onz,Cheung:2020gbf},  spin  \cite{Arkani-Hamed:2017jhn,Chung:2018kqs,Vines:2018gqi,Bern:2020buy,Guevara:2018wpp} and radiation effects \cite{Amati:1990xe, DiVecchia:2019myk, DiVecchia:2019kta, Bern:2020gjj, DiVecchia:2020ymx, Huber:2020xny,Damour:2020tta,DiVecchia:2021ndb}, and have spurred other new interesting  results (see e.g.~\cite{Foffa:2019hrb,Bini:2020uiq,Bini:2020rzn} for an incomplete list). 

The culminating product of the scattering amplitude program is the recent derivation of  the 4PM two-body Hamiltonian  \cite{Bern:2021dqo}.  At this order, a tail effect  is present \cite{Bini:2017wfr,Bini:2020hmy,Blanchet:2019rjs} and manifests an infrared divergence proportional to the leading-order ($G^3$) energy of the radiated   Bremsstrahlung, the gravitational waves emitted during the scattering of two masses approaching each other from infinity. Studies on the  leading-order gravitational Bremsstrahlung include \cite{Peters:1970mx,Thorne:1975aa,Crowley:1977us,Kovacs:1977uw,Kovacs:1978eu,Turner:1978zz,Westpfahl:1985tsl}. The full leading-order energy spectrum  found in \cite{Bern:2021dqo}
was independently obtained  in \cite{Herrmann:2021lqe} using the formalism of \cite{Kosower:2018adc}, which  derives classical observables from scattering amplitudes and their unitarity cuts. 

In this paper we study the gravitational Bremsstrahlung using a worldline approach inspired by  Non-Relativistic-General-Relativity (NRGR) \cite{Goldberger:2004jt} (see \cite{Goldberger:2007hy,Foffa:2013qca,Rothstein:2014sra,Porto:2016pyg,Levi:2018nxp} for reviews) 
and recently applied to the PM expansion \cite{Foffa:2013gja,Goldberger:2016iau,Goldberger:2017vcg,Kalin:2020mvi,Kalin:2020fhe}.  In particular, we first define the Feynman rules that allow us to derive the leading and next-to-leading order stress-energy tensor linearly coupled to gravity. From this we  compute the classical probability amplitude of graviton emission, which is directly related to the waveform in Fourier space. The amplitude is the basic ingredient for the computation of observables such as the radiated four-momentum and angular momentum, which we discuss in various limits and compare to the literature. 

Another article \cite{Jakobsen:2021smu}, whose content overlaps with ours, appeared while finalizing this work.

\section{Post-Minkowskian Effective field theory}

We consider the scattering of two gravitationally interacting  spinless bodies with mass $m_1$ and $m_2$ approaching each other from infinity.
The gravitational dynamics  is described by the usual Einstein-Hilbert action. Neglecting finite size effects, which would contribute at higher order in $G$ (see e.g.~\cite{Kalin:2020mvi, Kalin:2020lmz}), the bodies are treated as external sources described by point-particle actions.  We use the Polyakov-like parametrization of the action and fix the vielbein to unity.  This has the advantage of simplifying the gravitational coupling to the matter sources \cite{Galley:2013eba,Kuntz:2020gan,Kalin:2020mvi}. 
Therefore, using the mostly minus metric signature, setting $\hbar=c=1$ and defining the Planck mass as $\mpl \equiv  1/\sqrt{32 \pi G}$, we have
\be
\begin{split}
\label{eq:SAction}
S = &  - 2 \mpl^2 \int d^4 x \sqrt{-g} R  \\
& - \sum_{a=1,2} \frac{m_a}{2}\int d\tau_a  \big[ g_{\mu\nu}(x_a) \uu_a^\mu(\tau_a) \uu_a^\nu(\tau_a) + 1 \big] \; ,
\end{split}
\ee
where, for each body $a$, $\tau_a$ is its proper time 
and $\uu_a^\mu \equiv d x_a^\mu/d \tau_a$ is its four-velocity. 

To compute the waveform  we need the (pseudo) stress-energy tensor $T^{\mu\nu}$, defined as the linear term sourcing the gravitational field  in the effective action \cite{DeWitt:1967ub,Abbott:1981ke,Goldberger:2004jt}, i.e.,
\be
\label{bfea}
\Gamma [x_a, h_{\mu \nu} ] = - \frac{1}{2 \mpl} \int d^4 x T^{\mu \nu} (x) h_{\mu \nu} (x) \;.
\ee
In this equation $h_{\mu\nu} \equiv \mpl (g_{\mu\nu} - \eta_{\mu\nu})$   denotes a radiated field propagating on-shell, while $T^{\mu\nu}$ must include the contribution of both potential modes, i.e.~off-shell modes responsible for the conservative forces in the two-body system, and radiation modes. (We will come back to this split below.)

From the Fourier transform of $T^{\mu\nu}$, defined by
$ \tilde T^{\mu \nu} (k) = \int d^4 x  \,  T^{\mu\nu}(x) e^{ i k \cdot x}$,
one can compute the (classical) probability amplitude of one graviton emission with momentum $\vk$ and helicity $\lambda = \pm 2$ \cite{Goldberger:2004jt},
\be
\label{amplitude}
i {\cal A}_\lambda( k) = -  \frac{i}{2 \mpl} \epsilon^{ * \lambda}_{\mu \nu} (\vk)  \tilde T^{\mu \nu}(k) \;,
\ee 
where $\epsilon_{\mu \nu}^\lambda (\vk)$ is the transverse-traceless helicity-2 polarization tensor, with normalization $\epsilon^{* \lambda}_{\mu \nu} (\vk) \epsilon_{\lambda'}^{\mu \nu} (\vk) = \delta^{\lambda}_{ \lambda'}$ (see definition in App.~\ref{AppC}). At distances $r$ much larger than the interaction region, the waveform  is given in terms of the amplitude as  (see e.g.~\cite{Maggiore:1900zz})
\be
\label{waveform}
h_{\mu \nu} (x) = - \frac1{4 \pi r} \sum_{\lambda = \pm 2} \int \frac{d k^0}{2 \pi} e^{-i k^0 u} \epsilon_{\mu \nu}^\lambda (\vk) {\cal A}_\lambda (k) |_{k^\mu = k^0 n^\mu} \;, 
\ee
where $u \equiv t -r$. The amplitude is evaluated on-shell, i.e.~$k^\mu = k^0 n^\mu$, with $n^\mu \equiv (1, \vn)$ and $\vn$ the unitary vector pointing along the graviton trajectory.

We can obtain the stress-energy tensor defined above by matching eq.~\eqref{bfea} to the effective action computed order by order in $G$ using Feynman diagrams.
Let us now introduce  the Feynman rules. Adding the usual de Donder gauge-fixing term to eq. (\ref{eq:SAction}),
\be 
S_{\rm gf} = \int d^4 x \left[ \frac{1}{2}\partial_\rho h_{\mu\nu}\partial^\rho h^{\mu\nu}-\frac{1}{4}\partial_\rho h \partial^\rho h \right] \; ,
\ee
where $h \equiv \eta^{\mu\nu}h_{\mu\nu}$, from the quadratic part of the gravitational action one can extract the graviton propagator, 
\be 
\begin{tikzpicture}[baseline]
\begin{feynman}
\vertex [dot, label=90:$\mu\nu$] (a) {};
\vertex [dot, right=1.3cm of a, label=90:$\rho\sigma$] (b) {};
\diagram* {
(a) -- [gluon, edge label=$k$] (b),
}; 
\end{feynman}
\end{tikzpicture} = \frac{i}{k^2} P_{\mu\nu ; \rho\sigma}  \; ,
\ee
where $P_{\mu\nu ; \rho\sigma} \equiv \frac12 (\eta_{\mu\rho}\eta_{\nu\sigma}+
\eta_{\mu\sigma}\eta_{\nu\rho}-\eta_{\mu\nu}\eta_{\rho\sigma})$. As usual, we must specify the contour of integration in the complex $k^0$ plane by suitable boundary conditions. This is customarily done by splitting the gravitons into potential and radiation modes (see e.g.~\cite{Goldberger:2004jt,Kalin:2020mvi}). 
Potential modes never hit the pole $k^2=0$, so the choice of boundary conditions does not affect the calculations. 
For radiation modes one must impose retarded boundary conditions, i.e.~$[(k^0+i \epsilon)^2 - |\vk|^2]^{-1}$, to account only for outgoing gravitons.
Even though they are not relevant at the order in $G$ at which we work here, in general one must  treat with care the pole of  radiation modes 
  since they play a key role for hereditary effects at higher orders \cite{Goldberger:2009qd}. 
 
Finally, from the gravitational action one can derive the cubic interaction vertex, which is the only one relevant for this paper. In the de Donder gauge it can be found, for instance, in \cite{DeWitt:1967ub,Paszko:2010zz}.

Thanks to the Polyakov-like form, the point-particle action   contains only a linear interaction vertex. However, in order to isolate the powers of $G$, we parametrize the worldline by expanding around straight trajectories  \cite{Kalin:2020mvi, Kalin:2020fhe}, i.e., 
\begin{align}
x_a^\mu(\tau_a) & = b^\mu_a + u^\mu_a \tau_a + \delta^{(1)} x_a^\mu(\tau_a) +\dots \; ,\\
\uu_a^\mu (\tau_a) & = u_a^{\mu} + \delta^{(1)} u_a^\mu (\tau_a) +\dots \; .
\end{align}
Here $ u_a$ is the (constant) asymptotic incoming velocity and $b_a$ is the body displacement orthogonal to  it, $b_a \cdot u_a =0$, while  $\delta^{(1)} x_a^\mu$ and $ \delta^{(1)} u_a^\mu$ are respectively the deviation from the straight trajectory and constant velocity of body $a$ at order $G$, induced by the gravitational interaction.
Moreover, we define the impact parameter as $b^\mu \equiv b^\mu_1 - b^\mu_2$ and the relative Lorentz factor as
\be
\gamma \equiv u_1 \cdot u_2 = \frac{1}{\sqrt{1-v^2}} \;,
\ee
where $v$ is the relativistic relative velocity between the two bodies. 

The expansion of the worldline action in the second line of eq.~\eqref{eq:SAction} generates  two Feynman interaction rules that differ by their order in $G$. At zeroth order, we have (with $\int_q \equiv \int \frac{d^4 q}{(2 \pi)^4}$)
\begin{align}
\begin{tikzpicture}[baseline]
\begin{feynman}
\vertex [dot, label=90:$\tau_a$] (a) {};
\vertex [right=1.1cm of a] (b);
\diagram* {
(a) -- [gluon] (b),
}; 
\end{feynman}
\end{tikzpicture} & = -\frac{i m_a}{2\mpl} u^\mu_a u^\nu_a \int d\tau_a \int_q e^{-i q \cdot (b_a+u_a \tau_a)}  \; ,
\end{align}
where a filled dot  denotes the point particle evaluated using the straight worldline. At first order in $G$ we have
\be
\begin{split}
\begin{tikzpicture}[baseline]
\begin{feynman}
\vertex [empty dot, label=90:$\tau_a$] (a) {1};
\vertex [right=1.1cm of a] (b);
\diagram* {
(a) -- [gluon] (b),
}; 
\end{feynman}
\end{tikzpicture} & = -\frac{i m_a}{2\mpl} \int d\tau_a \int_q e^{-i q \cdot (b_a+u_a \tau_a)} \\
& \!\!\!\!\!\! \times \left( 2 \delta^{(1)} u_a^{(\mu}(\tau_a) u_a^{\nu)} - i ( q \cdot  \delta^{(1)} x_a (\tau_a) ) u_a^\mu u_a^\nu \right)  \; ,
\end{split}
\ee
where the correction  ${\cal O}(G^n)$ to the trajectory is denoted by the order $n$  inside the circle. Following \cite{Kalin:2020mvi}, the ${\cal O}(G)$ correction to the velocity and the trajectory can be computed by solving the geodesic equation obtained from the effective Lagrangian at order $G$. In the de Donder gauge  it reads, for particle $1$,
\begin{align}
\delta^{(1)} u_1^\mu (\tau)= & \    \frac{m_2}{4 \mpl^2}\int_q \dd(q \cdot u_2) \frac{e^{-i q\cdot b - i q\cdot u_1 \tau}}{q^2} B_1^\mu \;,  \\
\delta^{(1)} x_1^\mu (\tau)= & \ \frac{i m_2}{4 \mpl^2}\int_q \dd(q \cdot u_2) \frac{e^{-i q\cdot b - i q\cdot u_1 \tau}}{q^2 (q\cdot u_1 + i\varepsilon)} B_1^\mu \;,
\label{deltax}
\end{align}
where $B_1^\mu \equiv \frac{2\gamma^2-1}{2}\frac{q^\mu}{q\cdot u_1 + i\varepsilon}-2\gamma u_2^\mu +u_1^\mu$ and we use the notation $\dd^{(n)} (x)  \equiv (2 \pi)^n \delta^{(n)} (x)$. (An analogous expression holds for particle 2.) The $+i \epsilon$ in the above equations ensures to recover  straight motion in the asymptotic past, i.e.~$\delta^{(1)} u_1^\mu (- \infty)=0$ and $\delta^{(1)} x_1^\mu (- \infty)=0$. At our order in $G$, the deflected trajectories are completely determined  by  potential gravitons but in general one must take into account also radiation modes with appropriate boundary conditions.
Note also that at higher order it can be convenient to use different gauge-fixing conditions to simplify the graviton vertices \cite{Kalin:2020mvi}.

\section{Stress-energy tensor}

\begin{figure}[t!!]
\includegraphics[width=8cm]{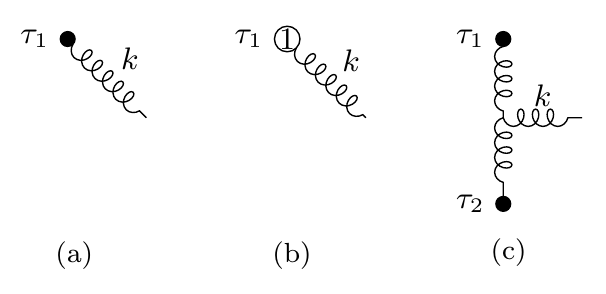}
\caption{The three Feynman diagrams needed for the computation of the stress-energy tensor up to NLO order in $G$. To compute the symmetric one, it is enough to exchange $1 \leftrightarrow 2 $.}
\label{Figure}
\end{figure}
The radiated field can be computed  in powers of $G$ in terms of the diagrams shown in Fig.~\ref{Figure}. 
The leading stress-energy tensor is obtained from Fig. 1a and corresponds to the one of  free point-particles, i.e.,  
 \be
 \tilde T^{\mu \nu}_{\text{{Fig.} 1a}}  (k)  = \sum_a m_a u_a^\mu u_a^\nu    e^{i k \cdot b_a}  \dd (\omega_a) \;,
 \label{SE1}
  \ee
 where  for convenience we define
 \be
 \omega_a \equiv k \cdot u_a \;, \qquad a = 1,2 \;.
 \ee
This generates a static and non-radiating contribution to the amplitude, proportional to $\dd (\omega_a) $. While this contribution can be neglected when computing the radiated momentum, it must be crucially included  for the computation of the angular momentum, as shown below.
 
 At the next order we find  
\begin{widetext}
 \begin{align}
 \tilde T^{\mu \nu}_{\text{Fig. 1b}} (k) =   & \ \frac{m_1 m_2}{4 \mpl^2} \int_{q_1,q_2} \mu_{1,2}(k)   \frac1{q_2^2} \bigg[  \frac{2 \gamma^2 - 1}{\omega_1 + i \epsilon} q_2^{(\mu} u_1^{\nu)} - 4 \gamma u_2^{(\mu} u_1^{\nu)}  -  \left( \frac{2 \gamma^2-1}{2} \frac{k \cdot q_2}{(\omega_1 + i \epsilon)^2}  - \frac{2 \gamma \omega_2}{\omega_1 + i \epsilon} -1 \right) u_1^\mu u_1^\nu  \bigg]  
\label{SE2} \;, \\
 \tilde T^{\mu \nu}_{\text{Fig. 1c}} (k) = & \ \frac{m_1 m_2}{4 \mpl^2   } \int_{q_1,q_2} \mu_{1,2}(k)   \frac{1}{q_1^2 q_2^2} \bigg[ \frac{2 \gamma^2 - 1}{2} q_2^\mu q_2^\nu + \left( 2  \omega_2^2  - q_1^2 \right) u_1^\mu u_1^\nu + 4 \gamma \omega_2 q_2^{(\mu} u_1^{\nu)}  \nonumber \\
&-\eta^{\mu \nu} \left( \gamma  \omega_1 \omega_2 + \frac{2 \gamma^2 - 1}{4} q^2_2 \right)+ 2 \left( \gamma q_1^2 -  \omega_1 \omega_2 \right) u_1^{(\mu} u_2^{\nu)} \bigg] \;,
\label{SE3}
 \end{align}
\end{widetext}
where 
\be
\mu_{1,2} (k) \equiv e^{i (q_1 \cdot b_1 + q_2 \cdot b_2)} \dd^{(4)}(k - q_1 -q_2) \dd(q_1 \cdot u_1) \dd(q_2 \cdot u_2) \;,
\ee
and we have used momentum conservation, on-shell and harmonic-gauge  conditions to simplify the final expression. 
Of course, we must also include the analogous diagrams with bodies $1$ and $2$ exchanged. The contribution  in Fig. 1b comes from evaluating the worldline along deflected trajectories while the  one  in Fig. 1c comes from the gravitational cubic interaction.  
We have checked that the sum of these two contributions is transverse for on-shell momenta, i.e.~$k_\mu \tilde T^{\mu \nu} =0$ for $k^2=0$, as expected for radiated gravitons. We have also verified that the finite part of the stress-energy tensor agrees with that computed in \cite{Goldberger:2016iau} once the contribution from the dilaton is removed.

\section{Amplitudes and waveforms}

 We expand the amplitude defined in eq.~\eqref{amplitude} in powers of $G$, ${\cal A}_\lambda =  {\cal A}^{(1)}_\lambda + {\cal A}^{(2)}_\lambda +  \ldots $.
Given the definition \eqref{amplitude} and the stress-energy tensor \eqref{SE1}, the leading order  reads
\be
{\cal A}^{(1)}_{\lambda} (k) = -  \frac{1}{2 \mpl} \sum_a m_a \; \epsilon^{*\lambda}_{\mu \nu } (\vn) u_a^\mu u_a^\nu \;   e^{i k \cdot b_a} \dd (\omega_a) \;.
\ee

The NLO can be obtained 
by summing  eqs.~\eqref{SE2} and \eqref{SE3} and inserting the result in eq.~\eqref{amplitude}. Integrating over one of the internal momenta,
\begin{widetext}
 \be
 \begin{split}
 {\cal A}_\lambda^{(2)} (k) & = -  \frac{m_1 m_2}{8 \mpl^3}  \epsilon_{\mu \nu}^{*\lambda} (\vn)  \bigg\{  e^{ i k \cdot b_1}  \bigg[
\left( - \frac{2 \gamma^2 -1}{2} \frac{k \cdot I_{(1)}}{(\omega_1+ i \epsilon)^2} + \frac{2 \gamma  \omega_2}{\omega_1+ i \epsilon}   I_{(0)}  +2  \omega _2^2 J_{(0)}  \right) u_1^\mu u_1^\nu  \\
&+ \left( \frac{2 \gamma^2 -1}{\omega_1+ i \epsilon} I_{(1)}^\mu   + 4 \gamma \omega_2 J_{(1)}^\mu  \right) u_1^\nu 
- 2 \left(   \gamma I_{(0)} +   \omega_1 \omega_2 J_{(0)} \right) u_1^\mu u_2^\nu + \frac{2 \gamma^2 -1}{2} J_{(2)}^{\mu \nu} \bigg]  
 \bigg\}  + (1 \leftrightarrow 2) \;,
 \label{eq:NLO_A_Int}
\end{split} 
 \ee
 \end{widetext}
where we have defined the following integrals,
\begin{align}
\label{eqs:1}
I^{\mu_1 \ldots \mu_n}_{(n)} & \equiv \int_q  \dd \left( q \cdot u_1 - \omega_1 \right) \dd  \left( q \cdot u_2 \right)\frac{e^{-i q \cdot b}}{q^2}  q^{\mu_1} \ldots q^{\mu_n} \; , \\
\label{eqs:2}
J^{\mu_1 \ldots \mu_n}_{(n)} & \equiv \int_q \dd  \left( q \cdot u_1 - \omega_1 \right) \dd \left( q \cdot u_2 \right)\frac{e^{-i q \cdot b}}{q^2 (k-q)^2} q^{\mu_1} \ldots q^{\mu_n} \; .
\end{align}
(The indices inside these integrals must be changed when evaluating the symmetric contribution $(1 \leftrightarrow 2)$.)
As detailed in App.~\ref{AppA}, the first set of integrals in eq.~\eqref{eqs:1} can be solved in terms of  Bessel functions. The second set of integrals  in eq.~\eqref{eqs:2} comes exclusively from the gravitational cubic interaction in Fig.~\ref{Figure}(c). Unfortunately we were not able to come up with an explicitly solution to these integrals. However, we can express them as  one- dimensional integrals over a Feynman parameter, involving Bessel functions. 

To simplify the treatment, from now on we choose a frame in which one of the two bodies,  say 2, is at rest.
Moreover, for convenience we can set $b_2^\mu = 0$ and $b_1^\mu = b^\mu$ and define the unit spatial vectors in the direction of $\vv$ and of the impact parameter $\vb$, respectively $ \ve_v \equiv {\vv}/{v}$ and $\ve_b = \vb/{|\vb|}$, with $ \ve_v \cdot  \ve_b =0$. We also define $v^\mu \equiv (1, v \ve_v)$ so that 
\be
\label{frame2}
u_2^\mu = \delta^\mu_0 \;, \qquad u_1^\mu =  \gamma v^\mu =  \gamma (1 , v \ve_v)  \; .
\ee
The energies of the radiated gravitons measured by the two bodies become, respectively, $\omega_2 =  k^0 \equiv \omega $ and $\omega_1 = \gamma \omega \, n \cdot v$. The amplitude simplifies to the following compact forms 
\begin{align}
{\cal A}^{(1)}_{\lambda} (k) &= -  \frac{m_1}{2 \mpl}    \; \frac{\gamma v^2}{ n \cdot v} \epsilon^{*\lambda}_{ ij  }  \ve_v^i \ve_v^j \;   \dd ( \omega ) e^{i k \cdot b} \;, \\
 {\cal A}_\lambda^{(2)} (k)  &= - \frac{ G m_1 m_2}{ \mpl \gamma v  }  \epsilon^{*\lambda}_{ij}    \ve_I^i \ve_J^j    { A}_{IJ} (k) e^{i k \cdot b} \; ,
 \label{eq:NLO_A}
\end{align}
where the functions $A_{IJ}$ can be obtained after solving the integrals \eqref{eqs:1} and \eqref{eqs:2}. We find
\begin{align}
{ A}_{vv} & =   c_{1} K_0 \big( z ( n \cdot v) \big) +  i c_2  \Big[ K_1 \big( z ( n \cdot v) \big)    - i \pi \delta \big( z ( n \cdot v) \big)  \Big]  \notag \\
  + & \int_0^1  \! dy \, e^{i y z v \vn \cdot \ve_b }  \Big[     d_{1} (y) z  K_1 \big(  z f(y) \big)  + c_0 K_0 \big( z f(y) \big) \Big]   \;, \label{eqs:A1} \\
{ A}_{vb} &  =   i c_{0}  \Big[ K_1 \big( z ( n \cdot v ) \big)  - i   \pi \delta \big( z ( n \cdot v) \big)  \Big] \notag \\ 
 & \quad +   i \int_0^1 \! dy \, e^{i y z v \vn \cdot \ve_b } d_{2} (y)  z  K_0 \big( z f(y) \big)   \; , \label{eqs:A2} \\
{ A}_{bb}  & =   \int_0^1 \! dy \, e^{i y z v \vn \cdot \ve_b }      d_{0} (y) z K_1 \big( z f(y) \big)  \; , \label{eqs:A3}
\end{align}
where $K_0$ and $K_1$ are modified Bessel functions of the second kind and we have introduced 
\be 
z \equiv \frac{|\vb| \omega}{ v}  \;, 
\ee
and 
\be
f(y) \equiv \sqrt{ (1-y)^2 (n \cdot v)^2 + 2 y    (1-y) (n \cdot v) + y^2/\gamma^2   } \;.
\ee
The coefficients  $c_0$, $c_1$ and $c_2$ depend on $v$ and on the relative angles between the graviton direction and the basis $(\ve_v,\ve_b)$. Moreover, $d_0$, $d_1$ and $d_2$ depend also on the integration parameter $y$. Their explicit form is given in App.~\ref{AppB}. In eqs.~\eqref{eqs:A1} and \eqref{eqs:A2} we have also included the  non-radiating contribution proportional to a delta function,\footnote{To compute this contribution we have used this integral: \be \int_q \delta(q \cdot u_1) \delta(q \cdot u_2) \frac{e^{- i q \cdot b} q^\mu }{q^2} = \frac{b^\mu}{2 \pi \gamma v |\vb|^2} \;. 
\ee} which  may become relevant, for instance, when computing the radiated angular momentum at  NLO.

For small velocities we find agreement between our amplitude and the waveform in Fourier space of \cite{Kovacs:1978eu}.
In this limit $f(y) \to 1$, $e^{i y z v \vn \cdot \ve_b } \to 1$, $\gamma \to 1$, and thus\footnote{The signs in front of $K_0$ and $K_1$ of the last term of  eqs.~(2.9b) and (2.9c) of \cite{Kovacs:1978eu} are opposite to ours because of a different Fourier transform convention.}
\begin{align}
{ A}_{vv} &\xrightarrow[v\to 0]{}    z  K_1 (  z  )  + K_0( z  )    \;, \label{eqs:A1KT} \\
{ A}_{vb} &\xrightarrow[v\to 0]{}     - i  \left[ K_1 ( z  ) + z  K_0 ( z )  - i   \pi \delta ( z  )  \right]  \; , \label{eqs:A2KT} \\
{ A}_{bb}  &\xrightarrow[v\to 0]{}   - z K_1 ( z  )  \; . \label{eqs:A3KT}
\end{align}
We have also checked that we recover their amplitude in the forward and backward limit (i.e.~$\vn$ along the direction of $\ve_v$), for which $\vn \cdot \ve_b \to 0$ and the integral in $y$ can be solved exactly. 
The waveform can be computed by replacing the amplitude in eq.~\eqref{waveform} and integrating in $k^0$. We discuss this calculation in  App.~\ref{app:WF}.

\section{Radiated  four-momentum}

In terms of the asymptotic waveform,  the radiated four-momentum  at infinity ($r \to \infty$) is given by \cite{Kovacs:1978eu,Damour:2020tta}\footnote{We are using a different normalization of $h_{\mu \nu}$ with respect to these references, which explains the absence of the prefactor $(32 \pi G)^{-1}$.}
\be
P^{\mu}_{\rm rad} =  \int d \Omega \, du \, r^2 \, n^\mu \, \dot h_{ij}  \dot  h_{ij}  \; ,
\ee
where a dot denotes the derivative with respect to the retarded time $u$ and $d \Omega$ is the integration surface element.

Using eq.~\eqref{waveform} for the waveform, this can be expressed in a manifestly Lorentz-invariant way in terms of the amplitude \eqref{amplitude} as \cite{Goldberger:2016iau} 
\be
\label{Goldpmu}
P^{\mu}_{\rm rad} = \sum_\lambda \int_k \dd (k^2) \theta(k^0) k^\mu \left| {\cal A}_\lambda (k)_{\rm finite} \right|^2  \; ,
\ee
where $\theta$ is the Heaviside step function and  on the right-hand side we take only the finite part of the amplitude, excluding the terms proportional to a delta function that do not contribute to $\dot h_{ij}$. Thus, at leading order $ \left| {\cal A}_\lambda (k)_{\rm finite} \right|^2 =  | {\cal A}^{(2)}_\lambda (k)_{\rm finite}  |^2 + \ldots$ and hence the radiated four-momentum starts at order $G^3$
 
Since the modulo squared of the amplitude is symmetric under $\vk \rightarrow -\vk$ the four-momentum cannot depend on the spatial direction $b^\mu$. Moreover, the energy measured in the frame of one body is the same as the one measured in the  frame of the other one, hence  the final result must be proportional to $u_1^\mu + u_2^\mu$. 
Using eq.~\eqref{eq:NLO_A}, we can write it  as
\be 
\label{totalmom}
P^{\mu}_{\rm rad} = \frac{G^3 m_1^2 m_2^2}{|\vb|^3} \frac{u_1^\mu + u_2^\mu}{\gamma + 1}  {\cal E}(\gamma)+ {\cal O}(G^4) \; ,
\ee 
which confirms that at this order the result has homogeneous mass dependence and is thus fixed by the probe limit \cite{Kovacs:1978eu,Bini:2020hmy,Herrmann:2021lqe}.
The function ${\cal E}(\gamma)$ can be found by integrating over the phase space the modulo squared of the amplitude,  
\be 
\label{Ecal}
{\cal E}(\gamma) =  \int \!\! d \Omega \int_0^\infty  dz \frac{d{\cal E}}{dz d \Omega}(z, \Omega; \gamma) \; 
\ee
with
\be 
\label{Ecalspec}
\frac{d{\cal E}}{dz d \Omega} \equiv \frac{2 v z^2  }{\pi^2\gamma^2 } \sum_\lambda  \big| \epsilon^{*\lambda}_{ij}    \ve_I^i \ve_J^j    { A}_{IJ} (z,\Omega) \big|^2 \; .
\ee
A more explicit but long expression of this function is reported in App.~\ref{app:E}, see eq.~\eqref{diffen}.

Due to the involved structure of the $y$ integrals in eq.~(\ref{eq:NLO_A}), we were unable to  compute ${\cal E}$ explicitly. Nevertheless, we can first compute the integrals in $y$ in the $v \ll 1$ regime at any order. Then we can perform the phase-space integral expressing the angular dependence in a particular coordinate system.
We have computed the energy up to order ${\cal O}(v^8)$, obtaining
\be 
\label{calE}
\frac{{\cal E}}{\pi } = \frac{37}{15} v +\frac{2393}{840} v^3 + \frac{61703}{10080} v^5 + \frac{3131839}{354816} v^7 + {\cal O}(v^9) \; .
\ee
The radiated energy in center-of-mass frame, $P_{\rm rad} \cdot u_{\rm CoM}$, where  
\be
u^\mu_{\rm CoM} = \frac{m_1 u_1^\mu + m_2 u_2^\mu}{\sqrt{m_1^2 + m_2^2 +2 m_1 m_2 \gamma}} \;,
\ee
agrees with the 2PN results \cite{Kovacs:1978eu,Blanchet:1989cu,Bini:2020hmy} while eq.~\eqref{calE} matches the  expansion of the  fully relativistic result recently found in \cite{Herrmann:2021lqe}. This is  a  non-trivial check of our NLO amplitude (\ref{eq:NLO_A}).

As an extra check, we can compute  the leading-order energy spectrum  in the soft limit, which is obtained by considering only wavelengths  of the emitted gravitons  much larger than  the interaction region, i.e.~$|\vb| \omega/v \ll 1$. For $E_{\rm rad} \equiv P^0_{\rm rad}$ this is given by
\be
\label{Espectrum}
\frac{d E_{\rm rad}}{d \omega}\Big|_{\omega \to 0}  = \frac{1}{2(2\pi)^3} \sum_{\lambda}\int d \Omega    | \omega {\cal A}_\lambda (k)_{\omega \to 0} |^2  \;.
\ee
In this limit the amplitude at order $G^2$ receives contributions exclusively from the diagram in Fig. 1b, so it is not affected by the gravitational self-interactions.
From eqs.~\eqref{eq:NLO_A}--\eqref{eqs:A3}, it reads
\be
\label{softA}
i {\cal A}^{(2)}_\lambda (k)_{\omega \to 0} = \frac{G m_1 m_2}{\mpl  |\vb|} \frac{1 }{\gamma \omega n \cdot v} \epsilon^{ * \lambda}_{ij} (c_2 \ve_v^i \ve_v^j + 2 c_0 \ve_v^i \ve_b^i) \;.
\ee 
Integrating eq.~\eqref{Espectrum} over the angles by fixing some  angular coordinate system and introducing  the function 
${\cal I}(v) \equiv -\frac{16}{3} +\frac{2}{v^2}+\frac{2(3v^2-1)}{v^3} \text{arctanh}(v) $ \cite{Damour:2020tta},
we obtain 
\be
\frac{d E_{\rm rad}}{d \omega}\Big|_{\omega \to 0}   =\frac{4}{\pi}\frac{(2\gamma^2-1)^2}{\gamma^2 v^2}\frac{G^3 m_1^2m_2^2}{|\vb|^2}  \,  {\cal I}(v) + {\cal O}(G^4) \;,
\label{Eraddiff}
\ee
which agrees with   \cite{Smarr:1977fy,DiVecchia:2021ndb}. 
We will come back to this result below.

\section{Radiated angular momentum}

The angular momentum lost by the system is another interesting observable as it can be related to the correction to the scattering angle due to radiation reaction \cite{Damour:2020tta}.
In terms of the asymptotic waveform this is given by \cite{Thorne:1980ru,Damour:2020tta} 
\be
\label{AMdef}
J^{i}_{\rm rad} = {\epsilon^{ijk}}  \int d \Omega \, du \, r^2 \left( 2 h_{j l}  \dot  h_{lk} - x^j \partial_k h_{lm} \dot h_{lm}  \right) \; .
\ee
As pointed out in \cite{Damour:2020tta}, the waveform at order $G$ is static and can be pulled out of the time integration leaving  with the computation of the {gravitational wave memory} $\Delta h_{ij}  \equiv  \int_{-\infty}^{+\infty} \!\! d u \, \dot{h}_{ij} $. This can be related to the classical amplitude by eq.~\eqref{waveform},
\be
\begin{split}
\Delta h_{ij}  = \frac{i}{4\pi r} \sum_{\lambda } \int \frac{d\omega}{2\pi} \epsilon^\lambda_{ij}  \dd(\omega) \omega {\cal A}_\lambda (k)_{\omega \to 0} \; ,
\end{split}
\ee
where from the right-hand side  it is clear that only the soft limit     contributes to the gravitational wave memory.  Moreover, since  at this order the soft limit is uniquely determined by the diagram in Fig. 1b,  the radiated angular momentum does not depend on the gravitational self-interaction, confirming \cite{Damour:2020tta}.

To compute the radiated angular momentum, it is convenient to introduce a  system of polar coordinates where $\vn = (\sin \theta \cos \phi, \sin \theta \sin \phi, \cos \theta)$ and  an orthonormal frame tangent to the sphere,  with $\ve_\theta = (\cos \theta \cos \phi , \cos \theta \sin \phi , -\sin \theta )$ and $\ve_\phi = (-\sin \phi, \cos \phi, 0)$. To express eq.~\eqref{AMdef} in terms of the amplitudes, 
we can rewrite the angular dependence in the polarization tensors of the  first term inside the parenthesis using $  2 \varepsilon^{ijk} \epsilon_{jl}^{\lambda } \epsilon_{lk}^{* \lambda' } = - i \lambda n^i \delta^{\lambda \lambda'}$.  The second term  can be rewritten by noticing that $\epsilon^{ijk}x^j \partial_k= i \hat L^i$, where $\hat L^i$ is the usual  orbital angular momentum operator, expressed in terms of the angles and their derivatives (see App.~\ref{AppC}).
Using $\epsilon_{lm}^{* \lambda' } {\hat {\bf L}}  \epsilon_{lm}^{\lambda} =   \lambda \cot \theta {\bf e}_\theta \delta^{\lambda \lambda'} $, we obtain 
\be
\label{Jrad}
{\bf J}_\text{rad}  = \sum_{\lambda}\int \frac{d \Omega}{(4\pi)^2} \,  \omega  {\cal A}^{(2)}_\lambda{}^* (k)_{\omega \to 0} \, {\hat {\bf J}} \bar {\cal A}_{\lambda}^{(1)} +{\cal O} (G^3) \;,
\ee
where ${\hat {\bf J}} \equiv \lambda   (\vn + \cot \theta \ve_\theta)  +  {\hat {\bf L}} $ 
and we have introduced $\bar {\cal A}_{\lambda}^{(1)}$ as the leading-order amplitude striped off of the delta function, i.e.~defined by
\be 
{\cal A}^{(1)}_{\lambda} (k)  = \bar {\cal A}_{\lambda}^{(1)} \dd (\omega) e^{i k \cdot b} \;.
\ee
One can perform the angular integral in eq.~\eqref{Jrad} by aligning $\ve_v$ and $\ve_b$ along any (mutually orthogonal) directions and  eventually obtains
\be
{\bf J}_\text{rad} = \frac{2 (2 \gamma^2 -1)}{\gamma v} \frac{G^2 m_1 m_2  J }{|\vb|^2}    {\cal I}(v) (  \ve_b \times \ve_v) \; ,
\ee
where $J = m_1 \gamma v |\vb|$ is the angular momentum at infinity. This result agrees with \cite{Damour:2020tta}.

As noticed in \cite{DiVecchia:2021ndb}, from eqs.~\eqref{Eraddiff} and \eqref{Jrad} we observe an intriguing proportionality between the energy spectrum in the soft limit  and the total emitted angular momentum.
We leave a more thorough exploration of this result for the future.

\section{Conclusion}

We have studied the gravitational Bremsstrahlung using a worldline approach.  In particular, we have computed through the use of Feynman diagrams, expanding  perturbatively in $G$, the leading and next-to-leading order   classical probability amplitude of graviton emission and consequently  the waveform in Fourier space. 
The next-to-leading order amplitude receives two contributions: one from  the deviation from straight orbits, which can be expressed in terms of  modified Bessel functions of the second kind; another from the cubic gravitational self-interaction, which we could rewrite as one-dimensional integrals over a Feynman parameter  of modified Bessel functions.  
When comparison was possible, we found agreement with earlier calculations of the waveforms \cite{Peters:1970mx,Kovacs:1977uw} in different limits.  

We have used the amplitude to compute the leading-order radiated angular momentum, recovering the result of \cite{Damour:2020tta}. Moreover, we have computed the total emitted four-momentum expanded in small velocities up to order $v^8$ and we found agreement with the recent results of \cite{Bern:2021dqo,Herrmann:2021lqe}. Unfortunately we were not able  to reproduce their fully relativistic result, which we leave for the future.
Nevertheless, 
we have built the foundations for an alternative derivation of the recent results obtained with amplitude techniques.

Another interesting limit is for small gravitational wave frequencies, where  the amplitude does not receive contributions from the gravitational interaction. We have computed the soft energy spectrum recovering an intriguing relation with the emitted angular momentum \cite{DiVecchia:2021ndb}. 
Future directions include the study of spin and finite-size effects and a more thorough investigation of the relations between differential observables.

\section*{Acknowledgements} 

We thank Laura Bernard, Luc Blanchet, Alberto Nicolis, Federico Piazza, Leong Khim Wong, Pierre Vanhove  and Gabriele Veneziano for  insightful discussions and correspondence. Moreover, we would like to thank the anonymous referee for several suggestions that helped to improve the paper and the Galileo Galilei
Institute, Florence,  the organizers and participants of the Workshop ``Gravitational Scattering, Inspiral and Radiation'' for interesting discussions.
This work was partially supported by the CNES and by the Munich Institute for Astro- and Particle Physics (MIAPP) which is funded by the Deutsche Forschungsgemeinschaft (DFG, German Research Foundation) under Germany's Excellence Strategy -- EXC-2094 -- 390783311.

\appendix

\section{Angular dependence}
\label{AppC}

We can introduce the transverse-traceless helicity-2 tensors, normalized to unity,  in terms of the orthonormal frame tangent to the sphere,  $\ve_\theta = (\cos \theta \cos \phi , \cos \theta \sin \phi , -\sin \theta )$ and $\ve_\phi = (-\sin \phi, \cos \phi, 0)$, used in the main text. We define
\be 
\label{eq:elicityA}
\epsilon^\pm_i \equiv \frac{1}{\sqrt{2}}( \pm \ve^i_\theta + i \ve^i_\phi) \;, \qquad \epsilon^{\pm 2}_{i j} = \epsilon^\pm_i \epsilon^\pm_j \; .
\ee
We can  relate these tensors to the (real) plus and cross parametrization often used in the literature by
\begin{align}
\epsilon^{\rm plus}_{ij} = \epsilon^+_{ij} + \epsilon^-_{ij}  \; , & & 
\epsilon^{\rm cross}_{ij} & = -i \left( \epsilon^+_{ij} - \epsilon^-_{ij} \right) \; .
\end{align}

For convenience, here we also  explicitly report the expression of the (orbital) angular momentum operator in terms of the same polar coordinates,
\begin{align}
{\hat L}^x & = i (\sin \phi \partial_\theta + \cot \theta \cos \phi \partial_\phi) \; ,  \\
{\hat L}^y & = -i (\cos \phi \partial_\theta - \cot \theta \sin \phi \partial_\phi) \; , \\
{\hat L}^z & = - i \partial_\phi \; .
\end{align}

\section{Integrals}
\label{AppA}

To compute the integrals in eq.~\eqref{eqs:1} we first need the master integral $I_{(0)}$, which can be solved by  going to the frame of body $2$ as in eq.~\eqref{frame2} and by removing the delta functions by integrating in $q^0$ and in the spatial momentum  along  $\vv$. This leaves us with
\be
I_{(0)} =  - \frac{1}{\gamma v}\int \frac{d^2 \vq_\perp}{(2 \pi)^2} \frac{e^{i \vq_\perp \cdot \vb }}{|\vq_\perp|^2 + \frac{\omega_1^2}{\gamma^2 v^2}}  = - \frac{1}{2 \pi \gamma v} K_0 \left( \frac{|\vb | \omega_1}{\gamma v} \right) \; ,
\ee
where we can write $|\vb| = \sqrt{-b^2}$ in a Lorentz-invariant fashion. 

We  use this result  to compute the descendant integrals $I_{(n)}^{\mu_1 \ldots \mu_n}$ (see analogous examples in \cite{Kosower:2018adc}). For instance, by  the presence of $\delta (q \cdot u_2) $ in the integrand, $I_{(1)}^{\mu} $ can only be a sum of two pieces, one proportional to $b^\mu$ and another proportional to  $u_1^\mu - \gamma u_2^\mu$. The piece proportional to $b^\mu$ can be computed by taking the derivative of  $I_{(0)}$ with respect to $b^\mu$ and projecting it along $b^\mu$ with proper normalization. It is easy to see that the other piece is proportional to $I_{(0)}$ upon  projecting $I_{(1)}^{\mu} $ along $u_1^\mu$ and taking into account  the first delta function. 

To compute the integrals in eq.~\eqref{eqs:2}, we  can proceed analogously. Although we were not able to solve the master integral $J_{(0)}$ in close form, we can express it in terms of an integral over a Feynman parameter as

\be
\begin{split}
J_{(0)}   &= \int_0^1  \! dy \, e^{- i y k \cdot b} \int_q  \dd\left( q \cdot u_1 + ( y - 1 ) \omega_1  \right) \times \\
& \qquad \qquad \qquad \qquad  \dd \left( q \cdot u_2 + y \omega_2 \right) {e^{-i q \cdot b}}/{q^4 } \\
&=\frac{|\vb|^2 }{4\pi \gamma v} \int_0^1  \! dy \, e^{- i y k \cdot b}\frac{K_1\left( z f(y)\right)}{z f(y)}\;,
\end{split}
\ee

where the integral in $q$ has been solved similarly to $I_{(0)}$.

\section{Coefficients}
\label{AppB}

The coefficients in eqs.~\eqref{eqs:A1}, \eqref{eqs:A2} and \eqref{eqs:A3} are
\be
\begin{split}
c_0  = & \ 1- 2 \gamma^2 \;, \quad c_{1} =  - c_0 +\frac{3 - 2 \gamma^2}{n \cdot v} \;, \quad c_2  =  v c_{0}    \frac{  \vn \cdot \ve_b}{  n \cdot v} \;, \\
d_{0} (y) & = f(y) c_0  \;, \\
d_{1} (y) & =   v^2\frac{ 4 \gamma^2 (y-1) (n \cdot v) -  c_0 (y-1)^2  -2 y -1 }{f(y)}  - d_0 (y)\;, \\
d_{2} (y) & =  - 1 + (1-y)c_0 (n \cdot v -1)   \;.
\end{split}
\ee

\section{Waveform in direct space}
\label{app:WF}

In this paper 
we focus on computing the emitted energy and angular momentum,  obtained from the waveform in {\em Fourier space} or, equivalently,  the amplitude of graviton emission. In this appendix, which was added in v2 of the paper to address one of the reviewer's comments, we show how to find the waveform in {\em direct space} from the expression of our amplitude.

First we replace the NLO amplitude \eqref{eq:NLO_A} in eq.~\eqref{waveform}, we go in the rest frame of particle 2 and integrate over $q^0$, removing $\dd (q \cdot u_2)$. Then we can get rid of the other delta function by integrating over $k^0$, which leads to a three-dimensional integral over $\vq$. More explicitly, at order $G^2$ one can find  (with $h_{\lambda} \equiv \epsilon_\lambda^{* \mu \nu}h_{\mu \nu}  $)
\begin{align}
h_{\pm 2}^{(2) } 
 = \frac{m_1 m_2 G }{ 8  \mpl r}\int_{\vq}e^{i \vq \cdot \tilde{\vb}} &  \bigg[ \frac{q^i {\cal N}^i_{\pm}}{\vq^2 \left( \vq\cdot\ve_v-i \epsilon \right)} \notag \\
& + \frac{q^iq^j {\cal M}^{ij}_{\pm}}{\vq^2 \left( \vq^2 + \vq \cdot L \cdot \vq \right)}\bigg] \, ,
\label{eq:waveform_direct}
\end{align}
where \cite{Jakobsen:2021smu}\footnote{To compare these expressions with those in \cite{Jakobsen:2021smu} one must replace
$\ve_v  \to \hat{\ve}_1$, $ \ve_b  \to \hat{\ve}_2 $, $ \ve_\theta  \to \hat{\bm{\theta}} $, $  \ve_\phi  \to \hat{\bm{\phi}} $, $ n^\mu         \to \rho^\mu $, $  v^\mu \to v^\mu_2/ \gamma  $ and use eq.~\eqref{eq:elicityA}.}
\begin{align}
\label{eq:Npm}
{\cal N}^i_{\pm} & \equiv 4 \frac{\gamma v}{(n \cdot v)^2}(\bm{\epsilon}^{\pm}\cdot \ve_v)^2\left[(1+v^2) \, n^i - 4 v \, e^i_v\right] \notag \\
&\qquad + 8\frac{\gamma (1+v^2)}{n \cdot v}(\bm{\epsilon}^{\pm}\cdot \ve_v)\epsilon^{i}_{\pm} \;, \\
\label{eq:Mpm}
{\cal M}^{ij}_{\pm} & \equiv 16 \frac{\gamma v^4}{(n \cdot v)^3}(\bm{\epsilon}^{\pm}\cdot \ve_v)^2 \, e^i_v e^j_v + 8\frac{\gamma(1+v^2)}{n \cdot v}\, \epsilon_{\pm}^{i}\epsilon_{\pm}^{j} \notag \\
&\qquad -32 \frac{\gamma v^2}{(n\cdot v)}(\bm{\epsilon}^{\pm}\cdot \ve_v)\, e_v^{(i}\epsilon_{\pm}^{j)} \;,
\end{align}
and
we have introduced
\begin{align}
\tilde{\vb} \equiv \vb + \frac{v}{n \cdot v}(u + \vb \cdot \vn) \,, & & L^{ij} \equiv 2 \frac{v}{n \cdot v} \ve_v^{(i} \vn^{j)} \,.
\end{align}
The integrations in $\vq$ can be performed following \cite{Jakobsen:2021smu}. Eventually one finds an expression of the waveform equivalent to that of this reference, which agrees with \cite{Kovacs:1978eu}.

\section{Energy and spectral dependence}
\label{app:E}
The spectral and angular dependence of the radiated four-momentum is given by eq.~\eqref{Ecalspec}. Using the expressions for the functions $A_{IJ}$ in eqs.~\eqref{eqs:A1}--\eqref{eqs:A3} and summing over the helicities, we find
\begin{widetext}
\be
\begin{split}
\label{diffen} 
 \frac{\pi^2 \gamma^2 }{2 v z^2  } &\frac{d{\cal E}}{dz d \Omega}   =    2a_{vv}^2  c_1^2 K_0^2 \left(z ( n \cdot v)\right) + 2 a_{vv} \left[ 4 a_{bb} c_0^2 +c_2 (4 a_{vb} c_0 +a_{vv} c_2 ) \right] K_1^2\left(z ( n \cdot v)\right)   \\
&  + 4c_1  K_0\left(z ( n \cdot v)\right) \big[ (2 a_{vb}^2 - a_{vv} a_{bb})  I_0^{(c)}  + a_{vv}^2  I_1^{(c)} -  2 a_{vv} a_{vb}   I_2^{(s)}  \big]   \\
&   + 4 K_1\left(z ( n \cdot v)\right) \big[ \left( 2 a_{bb} a_{vb} c_0 +(2 a_{vb}^2 - a_{vv} a_{bb}) c_2 \right) I_0^{(s)} +  (2 a_{vv} a_{vb} c_0 +a_{vv}^2 c_2) I_1^{(s)}    + 2 a_{vv} (2 a_{bb} c_0 + a_{vb} c_2)  I_2^{(c)} \big]  \\
& + 2  a_{bb}^2 \big[ \big( I_0^{(c)}   \big)^2  + \big( I_0^{(s)}   \big)^2 \big]  + 2 a_{vv}^2  \big[ \big( I_1^{(c)} \big)^2 + \big(  I_1^{(s)}   \big)^2 \big]   + 8 a_{vv} a_{bb} \big[ \big( I_2^{(c)} \big)^2 + \big(  I_2^{(s)} \big)^2 \big]  \\
& + 4 (2 a_{vb}^2 - a_{vv} a_{bb} )  \big( I_0^{(c)}  I_1^{(c)}    +  I_0^{(s)}  I_1^{(s)}  \big)  + 8 a_{vb} I_2^{(c)}  \big( a_{bb}  I_0^{(s)} +  a_{vv} I_1^{(s)}  \big)  - 8 a_{vb} I_2^{(s)}   \big( a_{bb}  I_0^{(c)}  +  a_{vv} I_1^{(c)}  \big)  \;,
\end{split}
\ee
\end{widetext}
where we have defined $a_{IJ} \equiv    [(\ve_\theta \cdot \ve_{I}) (\ve_\theta \cdot \ve_{J})+ (\ve_\phi \cdot \ve_{I}) (\ve_\phi \cdot \ve_{J})]/2 $,\footnote{Choosing $\ve_v$ along $z$ and  $\ve_b$ along $x$ we have 
$a_{vv} = \sin^2 \theta /2$, $ a_{vb} = - \sin \theta \cos \theta \cos \phi /2 $,  $ a_{bb} = [\cos^2 \theta \cos^2 \phi+ \sin^2 \phi]/2 $.}, 
and the two sets of integrals,
\be
\begin{split}
I_i^{(s)} (z,\Omega) &\equiv \int_0^1 dy \sin(y z v \vn \cdot \ve_b) g_i(z,\Omega;y) \;, \\
I_i^{(c)} (z,\Omega) &\equiv \int_0^1 dy \cos(y z v \vn \cdot \ve_b) g_i(z,\Omega;y) \;,
\end{split}
\ee
with
\be
\begin{split}
g_0(z,\Omega;y) &\equiv d_0(y) z K_1\left( z f(y)\right) \;, \\
g_1(z,\Omega;y) & \equiv  c_0  K_0\left( z f(y)\right)  +  d_1(y)z K_1\left( z f(y)\right) \;, \\
g_2(z,\Omega;y) &\equiv d_2 (y) z  K_0\left( z f(y)\right)  \;.  
\end{split}
\ee
It is straightforward to integrate analytically over the polar angle, while we were not able to integrate over the azimuthal one. Because of its length, we prefer not to report the integrated expression here.


\bibliographystyle{utphys}
\bibliography{ref}

\providecommand{\href}[2]{#2}\begingroup\raggedright\begin{thebibliography}{100}

\bibitem{Abbott:2016blz}
{\bf Virgo, LIGO Scientific} Collaboration, B.~P. Abbott {\em et.~al.},
  ``{Observation of Gravitational Waves from a Binary Black Hole Merger},''
  {\em Phys. Rev. Lett.} {\bf 116} (2016), no.~6 061102,
  \href{http://arxiv.org/abs/1602.03837}{{\tt 1602.03837}}.

\bibitem{TheLIGOScientific:2017qsa}
{\bf Virgo, LIGO Scientific} Collaboration, B.~P. Abbott {\em et.~al.},
  ``{GW170817: Observation of Gravitational Waves from a Binary Neutron Star
  Inspiral},'' {\em Phys. Rev. Lett.} {\bf 119} (2017), no.~16 161101,
  \href{http://arxiv.org/abs/1710.05832}{{\tt 1710.05832}}.

\bibitem{Bertotti:1956pxu}
B.~Bertotti, ``{On gravitational motion},'' {\em Nuovo Cim.} {\bf 4} (1956),
  no.~4 898--906.

\bibitem{Bertotti:1960wuq}
B.~Bertotti and J.~Plebanski, ``{Theory of gravitational perturbations in the
  fast motion approximation},'' {\em Annals Phys.} {\bf 11} (1960), no.~2
  169--200.

\bibitem{Havas:1962zz}
P.~Havas and J.~N. Goldberg, ``{Lorentz-Invariant Equations of Motion of Point
  Masses in the General Theory of Relativity},'' {\em Phys. Rev.} {\bf 128}
  (1962) 398--414.

\bibitem{Westpfahl:1979gu}
K.~Westpfahl and M.~Goller, ``{GRAVITATIONAL SCATTERING OF TWO RELATIVISTIC
  PARTICLES IN POSTLINEAR APPROXIMATION},'' {\em Lett. Nuovo Cim.} {\bf 26}
  (1979) 573--576.

\bibitem{Portilla:1980uz}
M.~Portilla, ``{SCATTERING OF TWO GRAVITATING PARTICLES: CLASSICAL APPROACH},''
  {\em J. Phys. A} {\bf 13} (1980) 3677--3683.

\bibitem{Bel:1981be}
L.~Bel, T.~Damour, N.~Deruelle, J.~Ibanez, and J.~Martin,
  ``{Poincar\'e-invariant gravitational field and equations of motion of two
  pointlike objects: The postlinear approximation of general relativity},''
  {\em Gen. Rel. Grav.} {\bf 13} (1981) 963--1004.

\bibitem{Westpfahl:1985tsl}
K.~Westpfahl, ``{High-Speed Scattering of Charged and Uncharged Particles in
  General Relativity},'' {\em Fortsch. Phys.} {\bf 33} (1985), no.~8 417--493.

\bibitem{Damour:2016gwp}
T.~Damour, ``{Gravitational scattering, post-Minkowskian approximation and
  Effective One-Body theory},'' {\em Phys. Rev. D} {\bf 94} (2016), no.~10
  104015, \href{http://arxiv.org/abs/1609.00354}{{\tt 1609.00354}}.

\bibitem{PhysRevD.97.044038}
T.~Damour, ``High-energy gravitational scattering and the general relativistic
  two-body problem,'' {\em Phys. Rev. D} {\bf 97} (Feb, 2018) 044038.

\bibitem{Blanchet:2018yvb}
L.~Blanchet and A.~S. Fokas, ``{Equations of motion of self-gravitating
  $N$-body systems in the first post-Minkowskian approximation},'' {\em Phys.
  Rev. D} {\bf 98} (2018), no.~8 084005,
  \href{http://arxiv.org/abs/1806.08347}{{\tt 1806.08347}}.

\bibitem{Blanchet:2013haa}
L.~Blanchet, ``{Gravitational Radiation from Post-Newtonian Sources and
  Inspiralling Compact Binaries},'' {\em Living Rev. Rel.} {\bf 17} (2014) 2,
  \href{http://arxiv.org/abs/1310.1528}{{\tt 1310.1528}}.

\bibitem{Blanchet:2018hut}
L.~Blanchet, ``{Analytic Approximations in GR and Gravitational Waves},'' {\em
  Int. J. Mod. Phys. D} {\bf 28} (2019), no.~06 1930011,
  \href{http://arxiv.org/abs/1812.07490}{{\tt 1812.07490}}.

\bibitem{Antonelli:2019ytb}
A.~Antonelli, A.~Buonanno, J.~Steinhoff, M.~van~de Meent, and J.~Vines,
  ``{Energetics of two-body Hamiltonians in post-Minkowskian gravity},'' {\em
  Phys. Rev. D} {\bf 99} (2019), no.~10 104004,
  \href{http://arxiv.org/abs/1901.07102}{{\tt 1901.07102}}.

\bibitem{Damour:2019lcq}
T.~Damour, ``{Classical and quantum scattering in post-Minkowskian gravity},''
  {\em Phys. Rev. D} {\bf 102} (2020), no.~2 024060,
  \href{http://arxiv.org/abs/1912.02139}{{\tt 1912.02139}}.

\bibitem{Bern:2008qj}
Z.~Bern, J.~J.~M. Carrasco, and H.~Johansson, ``{New Relations for Gauge-Theory
  Amplitudes},'' {\em Phys. Rev. D} {\bf 78} (2008) 085011,
  \href{http://arxiv.org/abs/0805.3993}{{\tt 0805.3993}}.

\bibitem{Bern:2010ue}
Z.~Bern, J.~J.~M. Carrasco, and H.~Johansson, ``{Perturbative Quantum Gravity
  as a Double Copy of Gauge Theory},'' {\em Phys. Rev. Lett.} {\bf 105} (2010)
  061602, \href{http://arxiv.org/abs/1004.0476}{{\tt 1004.0476}}.

\bibitem{Bern:2019prr}
Z.~Bern, J.~J. Carrasco, M.~Chiodaroli, H.~Johansson, and R.~Roiban, ``{The
  Duality Between Color and Kinematics and its Applications},''
  \href{http://arxiv.org/abs/1909.01358}{{\tt 1909.01358}}.

\bibitem{Bern:1994zx}
Z.~Bern, L.~J. Dixon, D.~C. Dunbar, and D.~A. Kosower, ``{One loop n point
  gauge theory amplitudes, unitarity and collinear limits},'' {\em Nucl. Phys.
  B} {\bf 425} (1994) 217--260, \href{http://arxiv.org/abs/hep-ph/9403226}{{\tt
  hep-ph/9403226}}.

\bibitem{Bern:1994cg}
Z.~Bern, L.~J. Dixon, D.~C. Dunbar, and D.~A. Kosower, ``{Fusing gauge theory
  tree amplitudes into loop amplitudes},'' {\em Nucl. Phys. B} {\bf 435} (1995)
  59--101, \href{http://arxiv.org/abs/hep-ph/9409265}{{\tt hep-ph/9409265}}.

\bibitem{Britto:2004nc}
R.~Britto, F.~Cachazo, and B.~Feng, ``{Generalized unitarity and one-loop
  amplitudes in N=4 super-Yang-Mills},'' {\em Nucl. Phys. B} {\bf 725} (2005)
  275--305, \href{http://arxiv.org/abs/hep-th/0412103}{{\tt hep-th/0412103}}.

\bibitem{Neill:2013wsa}
D.~Neill and I.~Z. Rothstein, ``{Classical Space-Times from the S Matrix},''
  {\em Nucl. Phys. B} {\bf 877} (2013) 177--189,
  \href{http://arxiv.org/abs/1304.7263}{{\tt 1304.7263}}.

\bibitem{Bjerrum-Bohr:2013bxa}
N.~E.~J. Bjerrum-Bohr, J.~F. Donoghue, and P.~Vanhove, ``{On-shell Techniques
  and Universal Results in Quantum Gravity},'' {\em JHEP} {\bf 02} (2014) 111,
  \href{http://arxiv.org/abs/1309.0804}{{\tt 1309.0804}}.

\bibitem{Luna:2017dtq}
A.~Luna, I.~Nicholson, D.~O'Connell, and C.~D. White, ``{Inelastic Black Hole
  Scattering from Charged Scalar Amplitudes},'' {\em JHEP} {\bf 03} (2018) 044,
  \href{http://arxiv.org/abs/1711.03901}{{\tt 1711.03901}}.

\bibitem{Bjerrum-Bohr:2018xdl}
N.~E.~J. Bjerrum-Bohr, P.~H. Damgaard, G.~Festuccia, L.~Plant\'e, and
  P.~Vanhove, ``{General Relativity from Scattering Amplitudes},'' {\em Phys.
  Rev. Lett.} {\bf 121} (2018), no.~17 171601,
  \href{http://arxiv.org/abs/1806.04920}{{\tt 1806.04920}}.

\bibitem{Kosower:2018adc}
D.~A. Kosower, B.~Maybee, and D.~O'Connell, ``{Amplitudes, Observables, and
  Classical Scattering},'' {\em JHEP} {\bf 02} (2019) 137,
  \href{http://arxiv.org/abs/1811.10950}{{\tt 1811.10950}}.

\bibitem{Cheung:2018wkq}
C.~Cheung, I.~Z. Rothstein, and M.~P. Solon, ``{From Scattering Amplitudes to
  Classical Potentials in the Post-Minkowskian Expansion},'' {\em Phys. Rev.
  Lett.} {\bf 121} (2018), no.~25 251101,
  \href{http://arxiv.org/abs/1808.02489}{{\tt 1808.02489}}.

\bibitem{Cristofoli:2019neg}
A.~Cristofoli, N.~E.~J. Bjerrum-Bohr, P.~H. Damgaard, and P.~Vanhove,
  ``{Post-Minkowskian Hamiltonians in general relativity},'' {\em Phys. Rev. D}
  {\bf 100} (2019), no.~8 084040, \href{http://arxiv.org/abs/1906.01579}{{\tt
  1906.01579}}.

\bibitem{Cristofoli:2020uzm}
A.~Cristofoli, P.~H. Damgaard, P.~Di~Vecchia, and C.~Heissenberg,
  ``{Second-order Post-Minkowskian scattering in arbitrary dimensions},'' {\em
  JHEP} {\bf 07} (2020) 122, \href{http://arxiv.org/abs/2003.10274}{{\tt
  2003.10274}}.

\bibitem{Veltman:1975vx}
M.~J.~G. Veltman, ``{Quantum Theory of Gravitation},'' {\em Conf. Proc. C} {\bf
  7507281} (1975) 265--327.

\bibitem{DeWitt:1967yk}
B.~S. DeWitt, ``{Quantum Theory of Gravity. 1. The Canonical Theory},'' {\em
  Phys. Rev.} {\bf 160} (1967) 1113--1148.

\bibitem{DeWitt:1967ub}
B.~S. DeWitt, ``{Quantum Theory of Gravity. 2. The Manifestly Covariant
  Theory},'' {\em Phys. Rev.} {\bf 162} (1967) 1195--1239.

\bibitem{DeWitt:1967uc}
B.~S. DeWitt, ``{Quantum Theory of Gravity. 3. Applications of the Covariant
  Theory},'' {\em Phys. Rev.} {\bf 162} (1967) 1239--1256.

\bibitem{PhysRevD.50.3874}
J.~F. Donoghue, ``General relativity as an effective field theory: The leading
  quantum corrections,'' {\em Phys. Rev. D} {\bf 50} (Sep, 1994) 3874--3888.

\bibitem{PhysRevD.67.084033}
N.~E.~J. Bjerrum-Bohr, J.~F. Donoghue, and B.~R. Holstein, ``Quantum
  gravitational corrections to the nonrelativistic scattering potential of two
  masses,'' {\em Phys. Rev. D} {\bf 67} (Apr, 2003) 084033.

\bibitem{Iwasaki:1971iy}
Y.~Iwasaki, ``{Fourth-order gravitational potential based on quantum field
  theory},'' {\em Lett. Nuovo Cim.} {\bf 1} (1971) 783--786.

\bibitem{Iwasaki:1971vb}
Y.~Iwasaki, ``{Quantum theory of gravitation vs. classical theory. -
  fourth-order potential},'' {\em Prog. Theor. Phys.} {\bf 46} (1971)
  1587--1609.

\bibitem{Mougiakakos:2020laz}
S.~Mougiakakos and P.~Vanhove, ``{Schwarzschild-Tangherlini metric from
  scattering amplitudes in various dimensions},'' {\em Phys. Rev. D} {\bf 103}
  (2021), no.~2 026001, \href{http://arxiv.org/abs/2010.08882}{{\tt
  2010.08882}}.

\bibitem{Goldberger:2016iau}
W.~D. Goldberger and A.~K. Ridgway, ``{Radiation and the classical double copy
  for color charges},'' {\em Phys. Rev. D} {\bf 95} (2017), no.~12 125010,
  \href{http://arxiv.org/abs/1611.03493}{{\tt 1611.03493}}.

\bibitem{Goldberger:2017vcg}
W.~D. Goldberger and A.~K. Ridgway, ``{Bound states and the classical double
  copy},'' {\em Phys. Rev. D} {\bf 97} (2018), no.~8 085019,
  \href{http://arxiv.org/abs/1711.09493}{{\tt 1711.09493}}.

\bibitem{Kalin:2020mvi}
G.~K\"alin and R.~A. Porto, ``{Post-Minkowskian Effective Field Theory for
  Conservative Binary Dynamics},'' {\em JHEP} {\bf 11} (2020) 106,
  \href{http://arxiv.org/abs/2006.01184}{{\tt 2006.01184}}.

\bibitem{Loebbert:2020aos}
F.~Loebbert, J.~Plefka, C.~Shi, and T.~Wang, ``{Three-body effective potential
  in general relativity at second post-Minkowskian order and resulting
  post-Newtonian contributions},'' {\em Phys. Rev. D} {\bf 103} (2021), no.~6
  064010, \href{http://arxiv.org/abs/2012.14224}{{\tt 2012.14224}}.

\bibitem{Mogull:2020sak}
G.~Mogull, J.~Plefka, and J.~Steinhoff, ``{Classical black hole scattering from
  a worldline quantum field theory},'' {\em JHEP} {\bf 02} (2021) 048,
  \href{http://arxiv.org/abs/2010.02865}{{\tt 2010.02865}}.

\bibitem{Kalin:2019rwq}
G.~K\"alin and R.~A. Porto, ``{From Boundary Data to Bound States},'' {\em
  JHEP} {\bf 01} (2020) 072, \href{http://arxiv.org/abs/1910.03008}{{\tt
  1910.03008}}.

\bibitem{Kalin:2019inp}
G.~K\"alin and R.~A. Porto, ``{From boundary data to bound states. Part II.
  Scattering angle to dynamical invariants (with twist)},'' {\em JHEP} {\bf 02}
  (2020) 120, \href{http://arxiv.org/abs/1911.09130}{{\tt 1911.09130}}.

\bibitem{Bern:2019nnu}
Z.~Bern, C.~Cheung, R.~Roiban, C.-H. Shen, M.~P. Solon, and M.~Zeng,
  ``{Scattering Amplitudes and the Conservative Hamiltonian for Binary Systems
  at Third Post-Minkowskian Order},'' {\em Phys. Rev. Lett.} {\bf 122} (2019),
  no.~20 201603, \href{http://arxiv.org/abs/1901.04424}{{\tt 1901.04424}}.

\bibitem{Bern:2019crd}
Z.~Bern, C.~Cheung, R.~Roiban, C.-H. Shen, M.~P. Solon, and M.~Zeng, ``{Black
  Hole Binary Dynamics from the Double Copy and Effective Theory},'' {\em JHEP}
  {\bf 10} (2019) 206, \href{http://arxiv.org/abs/1908.01493}{{\tt
  1908.01493}}.

\bibitem{Cheung:2020gyp}
C.~Cheung and M.~P. Solon, ``{Classical gravitational scattering at $
  \mathcal{O} $(G$^{3}$) from Feynman diagrams},'' {\em JHEP} {\bf 06} (2020)
  144, \href{http://arxiv.org/abs/2003.08351}{{\tt 2003.08351}}.

\bibitem{Kalin:2020fhe}
G.~K\"alin, Z.~Liu, and R.~A. Porto, ``{Conservative Dynamics of Binary Systems
  to Third Post-Minkowskian Order from the Effective Field Theory Approach},''
  {\em Phys. Rev. Lett.} {\bf 125} (2020), no.~26 261103,
  \href{http://arxiv.org/abs/2007.04977}{{\tt 2007.04977}}.

\bibitem{Kalin:2020lmz}
G.~K\"alin, Z.~Liu, and R.~A. Porto, ``{Conservative Tidal Effects in Compact
  Binary Systems to Next-to-Leading Post-Minkowskian Order},'' {\em Phys. Rev.
  D} {\bf 102} (2020) 124025, \href{http://arxiv.org/abs/2008.06047}{{\tt
  2008.06047}}.

\bibitem{Bern:2020uwk}
Z.~Bern, J.~Parra-Martinez, R.~Roiban, E.~Sawyer, and C.-H. Shen, ``{Leading
  Nonlinear Tidal Effects and Scattering Amplitudes},''
  \href{http://arxiv.org/abs/2010.08559}{{\tt 2010.08559}}.

\bibitem{Cheung:2020sdj}
C.~Cheung and M.~P. Solon, ``{Tidal Effects in the Post-Minkowskian
  Expansion},'' {\em Phys. Rev. Lett.} {\bf 125} (2020), no.~19 191601,
  \href{http://arxiv.org/abs/2006.06665}{{\tt 2006.06665}}.

\bibitem{AccettulliHuber:2020oou}
M.~Accettulli~Huber, A.~Brandhuber, S.~De~Angelis, and G.~Travaglini,
  ``{Eikonal phase matrix, deflection angle and time delay in effective field
  theories of gravity},'' {\em Phys. Rev. D} {\bf 102} (2020), no.~4 046014,
  \href{http://arxiv.org/abs/2006.02375}{{\tt 2006.02375}}.

\bibitem{Haddad:2020que}
K.~Haddad and A.~Helset, ``{Tidal effects in quantum field theory},'' {\em
  JHEP} {\bf 12} (2020) 024, \href{http://arxiv.org/abs/2008.04920}{{\tt
  2008.04920}}.

\bibitem{Aoude:2020onz}
R.~Aoude, K.~Haddad, and A.~Helset, ``{On-shell heavy particle effective
  theories},'' {\em JHEP} {\bf 05} (2020) 051,
  \href{http://arxiv.org/abs/2001.09164}{{\tt 2001.09164}}.

\bibitem{Cheung:2020gbf}
C.~Cheung, N.~Shah, and M.~P. Solon, ``{Mining the Geodesic Equation for
  Scattering Data},'' {\em Phys. Rev. D} {\bf 103} (2021), no.~2 024030,
  \href{http://arxiv.org/abs/2010.08568}{{\tt 2010.08568}}.

\bibitem{Arkani-Hamed:2017jhn}
N.~Arkani-Hamed, T.-C. Huang, and Y.-t. Huang, ``{Scattering Amplitudes For All
  Masses and Spins},'' \href{http://arxiv.org/abs/1709.04891}{{\tt
  1709.04891}}.

\bibitem{Chung:2018kqs}
M.-Z. Chung, Y.-T. Huang, J.-W. Kim, and S.~Lee, ``{The simplest massive
  S-matrix: from minimal coupling to Black Holes},'' {\em JHEP} {\bf 04} (2019)
  156, \href{http://arxiv.org/abs/1812.08752}{{\tt 1812.08752}}.

\bibitem{Vines:2018gqi}
J.~Vines, J.~Steinhoff, and A.~Buonanno, ``{Spinning-black-hole scattering and
  the test-black-hole limit at second post-Minkowskian order},'' {\em Phys.
  Rev. D} {\bf 99} (2019), no.~6 064054,
  \href{http://arxiv.org/abs/1812.00956}{{\tt 1812.00956}}.

\bibitem{Bern:2020buy}
Z.~Bern, A.~Luna, R.~Roiban, C.-H. Shen, and M.~Zeng, ``{Spinning Black Hole
  Binary Dynamics, Scattering Amplitudes and Effective Field Theory},''
  \href{http://arxiv.org/abs/2005.03071}{{\tt 2005.03071}}.

\bibitem{Guevara:2018wpp}
A.~Guevara, A.~Ochirov, and J.~Vines, ``{Scattering of Spinning Black Holes
  from Exponentiated Soft Factors},'' {\em JHEP} {\bf 09} (2019) 056,
  \href{http://arxiv.org/abs/1812.06895}{{\tt 1812.06895}}.

\bibitem{Amati:1990xe}
D.~Amati, M.~Ciafaloni, and G.~Veneziano, ``{Higher Order Gravitational
  Deflection and Soft Bremsstrahlung in Planckian Energy Superstring
  Collisions},'' {\em Nucl. Phys. B} {\bf 347} (1990) 550--580.

\bibitem{DiVecchia:2019myk}
P.~Di~Vecchia, A.~Luna, S.~G. Naculich, R.~Russo, G.~Veneziano, and C.~D.
  White, ``{A tale of two exponentiations in ${\cal N}=8$ supergravity},'' {\em
  Phys. Lett. B} {\bf 798} (2019) 134927,
  \href{http://arxiv.org/abs/1908.05603}{{\tt 1908.05603}}.

\bibitem{DiVecchia:2019kta}
P.~Di~Vecchia, S.~G. Naculich, R.~Russo, G.~Veneziano, and C.~D. White, ``{A
  tale of two exponentiations in $ \mathcal{N} $ = 8 supergravity at subleading
  level},'' {\em JHEP} {\bf 03} (2020) 173,
  \href{http://arxiv.org/abs/1911.11716}{{\tt 1911.11716}}.

\bibitem{Bern:2020gjj}
Z.~Bern, H.~Ita, J.~Parra-Martinez, and M.~S. Ruf, ``{Universality in the
  classical limit of massless gravitational scattering},'' {\em Phys. Rev.
  Lett.} {\bf 125} (2020), no.~3 031601,
  \href{http://arxiv.org/abs/2002.02459}{{\tt 2002.02459}}.

\bibitem{DiVecchia:2020ymx}
P.~Di~Vecchia, C.~Heissenberg, R.~Russo, and G.~Veneziano, ``{Universality of
  ultra-relativistic gravitational scattering},'' {\em Phys. Lett. B} {\bf 811}
  (2020) 135924, \href{http://arxiv.org/abs/2008.12743}{{\tt 2008.12743}}.

\bibitem{Huber:2020xny}
M.~Accettulli~Huber, A.~Brandhuber, S.~De~Angelis, and G.~Travaglini, ``{From
  amplitudes to gravitational radiation with cubic interactions and tidal
  effects},'' {\em Phys. Rev. D} {\bf 103} (2021), no.~4 045015,
  \href{http://arxiv.org/abs/2012.06548}{{\tt 2012.06548}}.

\bibitem{Damour:2020tta}
T.~Damour, ``{Radiative contribution to classical gravitational scattering at
  the third order in $G$},'' {\em Phys. Rev. D} {\bf 102} (2020), no.~12
  124008, \href{http://arxiv.org/abs/2010.01641}{{\tt 2010.01641}}.

\bibitem{DiVecchia:2021ndb}
P.~Di~Vecchia, C.~Heissenberg, R.~Russo, and G.~Veneziano, ``{Radiation
  Reaction from Soft Theorems},'' \href{http://arxiv.org/abs/2101.05772}{{\tt
  2101.05772}}.

\bibitem{Foffa:2019hrb}
S.~Foffa, P.~Mastrolia, R.~Sturani, C.~Sturm, and W.~J. Torres~Bobadilla,
  ``{Static two-body potential at fifth post-Newtonian order},'' {\em Phys.
  Rev. Lett.} {\bf 122} (2019), no.~24 241605,
  \href{http://arxiv.org/abs/1902.10571}{{\tt 1902.10571}}.

\bibitem{Bini:2020uiq}
D.~Bini, T.~Damour, A.~Geralico, S.~Laporta, and P.~Mastrolia, ``{Gravitational
  dynamics at $O(G^6)$: perturbative gravitational scattering meets
  experimental mathematics},'' \href{http://arxiv.org/abs/2008.09389}{{\tt
  2008.09389}}.

\bibitem{Bini:2020rzn}
D.~Bini, T.~Damour, A.~Geralico, S.~Laporta, and P.~Mastrolia, ``{Gravitational
  scattering at the seventh order in $G$: nonlocal contribution at the sixth
  post-Newtonian accuracy},'' {\em Phys. Rev. D} {\bf 103} (2021), no.~4
  044038, \href{http://arxiv.org/abs/2012.12918}{{\tt 2012.12918}}.

\bibitem{Bern:2021dqo}
Z.~Bern, J.~Parra-Martinez, R.~Roiban, M.~S. Ruf, C.-H. Shen, M.~P. Solon, and
  M.~Zeng, ``{Scattering Amplitudes and Conservative Binary Dynamics at ${\cal
  O}(G^4)$},'' {\em Phys. Rev. Lett.} {\bf 126} (2021), no.~17 171601,
  \href{http://arxiv.org/abs/2101.07254}{{\tt 2101.07254}}.

\bibitem{Bini:2017wfr}
D.~Bini and T.~Damour, ``{Gravitational scattering of two black holes at the
  fourth post-Newtonian approximation},'' {\em Phys. Rev. D} {\bf 96} (2017),
  no.~6 064021, \href{http://arxiv.org/abs/1706.06877}{{\tt 1706.06877}}.

\bibitem{Bini:2020hmy}
D.~Bini, T.~Damour, and A.~Geralico, ``{Sixth post-Newtonian nonlocal-in-time
  dynamics of binary systems},'' {\em Phys. Rev. D} {\bf 102} (2020), no.~8
  084047, \href{http://arxiv.org/abs/2007.11239}{{\tt 2007.11239}}.

\bibitem{Blanchet:2019rjs}
L.~Blanchet, S.~Foffa, F.~Larrouturou, and R.~Sturani, ``{Logarithmic tail
  contributions to the energy function of circular compact binaries},'' {\em
  Phys. Rev. D} {\bf 101} (2020), no.~8 084045,
  \href{http://arxiv.org/abs/1912.12359}{{\tt 1912.12359}}.

\bibitem{Peters:1970mx}
P.~C. Peters, ``{Relativistic gravitational bremsstrahlung},'' {\em Phys. Rev.
  D} {\bf 1} (1970) 1559--1571.

\bibitem{Thorne:1975aa}
K.~S. Thorne and S.~J. Kovacs, ``{The Generation of Gravitational Waves. 1.
  Weak-field sources},'' {\em Astrophys. J.} {\bf 200} (1975) 245--262.

\bibitem{Crowley:1977us}
R.~J. Crowley and K.~S. Thorne, ``{The Generation of Gravitational Waves. 2.
  The Postlinear Formalism Revisited},'' {\em Astrophys. J.} {\bf 215} (1977)
  624--635.

\bibitem{Kovacs:1977uw}
S.~Kovacs and K.~Thorne, ``{The Generation of Gravitational Waves. 3.
  Derivation of Bremsstrahlung Formulas},'' {\em Astrophys. J.} {\bf 217}
  (1977) 252--280.

\bibitem{Kovacs:1978eu}
S.~Kovacs and K.~Thorne, ``{The Generation of Gravitational Waves. 4.
  Bremsstrahlung},'' {\em Astrophys. J.} {\bf 224} (1978) 62--85.

\bibitem{Turner:1978zz}
M.~Turner and C.~M. Will, ``{Post-Newtonian gravitational bremsstrahlung},''
  {\em Astrophys. J.} {\bf 220} (1978) 1107--1124.

\bibitem{Herrmann:2021lqe}
E.~Herrmann, J.~Parra-Martinez, M.~S. Ruf, and M.~Zeng, ``{Gravitational
  Bremsstrahlung from Reverse Unitarity},'' {\em Phys. Rev. Lett.} {\bf 126}
  (2021), no.~20 201602, \href{http://arxiv.org/abs/2101.07255}{{\tt
  2101.07255}}.

\bibitem{Goldberger:2004jt}
W.~D. Goldberger and I.~Z. Rothstein, ``{An Effective field theory of gravity
  for extended objects},'' {\em Phys. Rev. D} {\bf 73} (2006) 104029,
  \href{http://arxiv.org/abs/hep-th/0409156}{{\tt hep-th/0409156}}.

\bibitem{Goldberger:2007hy}
W.~D. Goldberger, ``{Les Houches lectures on effective field theories and
  gravitational radiation},'' in {\em {Les Houches Summer School - Session 86:
  Particle Physics and Cosmology: The Fabric of Spacetime}}, 1, 2007.
\newblock \href{http://arxiv.org/abs/hep-ph/0701129}{{\tt hep-ph/0701129}}.

\bibitem{Foffa:2013qca}
S.~Foffa and R.~Sturani, ``{Effective field theory methods to model compact
  binaries},'' {\em Class. Quant. Grav.} {\bf 31} (2014), no.~4 043001,
  \href{http://arxiv.org/abs/1309.3474}{{\tt 1309.3474}}.

\bibitem{Rothstein:2014sra}
I.~Z. Rothstein, ``{Progress in effective field theory approach to the binary
  inspiral problem},'' {\em Gen. Rel. Grav.} {\bf 46} (2014) 1726.

\bibitem{Porto:2016pyg}
R.~A. Porto, ``{The effective field theorist\textquoteright{}s approach to
  gravitational dynamics},'' {\em Phys. Rept.} {\bf 633} (2016) 1--104,
  \href{http://arxiv.org/abs/1601.04914}{{\tt 1601.04914}}.

\bibitem{Levi:2018nxp}
M.~Levi, ``{Effective Field Theories of Post-Newtonian Gravity: A comprehensive
  review},'' {\em Rept. Prog. Phys.} {\bf 83} (2020), no.~7 075901,
  \href{http://arxiv.org/abs/1807.01699}{{\tt 1807.01699}}.

\bibitem{Foffa:2013gja}
S.~Foffa, ``{Gravitating binaries at 5PN in the post-Minkowskian
  approximation},'' {\em Phys. Rev. D} {\bf 89} (2014), no.~2 024019,
  \href{http://arxiv.org/abs/1309.3956}{{\tt 1309.3956}}.

\bibitem{Jakobsen:2021smu}
G.~U. Jakobsen, G.~Mogull, J.~Plefka, and J.~Steinhoff, ``{Classical
  Gravitational Bremsstrahlung from a Worldline Quantum Field Theory},'' {\em
  Phys. Rev. Lett.} {\bf 126} (2021), no.~20 201103,
  \href{http://arxiv.org/abs/2101.12688}{{\tt 2101.12688}}.

\bibitem{Galley:2013eba}
C.~R. Galley and R.~A. Porto, ``{Gravitational self-force in the
  ultra-relativistic limit: the ''large-$N$'' expansion},'' {\em JHEP} {\bf 11}
  (2013) 096, \href{http://arxiv.org/abs/1302.4486}{{\tt 1302.4486}}.

\bibitem{Kuntz:2020gan}
A.~Kuntz, ``{Half-solution to the two-body problem in General Relativity},''
  {\em Phys. Rev. D} {\bf 102} (2020), no.~6 064019,
  \href{http://arxiv.org/abs/2003.03366}{{\tt 2003.03366}}.

\bibitem{Abbott:1981ke}
L.~Abbott, ``{Introduction to the Background Field Method},'' {\em Acta Phys.
  Polon. B} {\bf 13} (1982) 33.

\bibitem{Maggiore:1900zz}
M.~Maggiore, {\em {Gravitational Waves. Vol. 1: Theory and Experiments}}.
\newblock Oxford Master Series in Physics. Oxford University Press, 2007.

\bibitem{Goldberger:2009qd}
W.~D. Goldberger and A.~Ross, ``{Gravitational radiative corrections from
  effective field theory},'' {\em Phys. Rev. D} {\bf 81} (2010) 124015,
  \href{http://arxiv.org/abs/0912.4254}{{\tt 0912.4254}}.

\bibitem{Paszko:2010zz}
R.~Paszko and A.~Accioly, ``{Equivalence between the semiclassical and
  effective approaches to gravity},'' {\em Class. Quant. Grav.} {\bf 27} (2010)
  145012.

\bibitem{Blanchet:1989cu}
L.~Blanchet and G.~Schaefer, ``{Higher order gravitational radiation losses in
  binary systems},'' {\em Mon. Not. Roy. Astron. Soc.} {\bf 239} (1989)
  845--867. [Erratum: Mon.Not.Roy.Astron.Soc. 242, 704 (1990)].

\bibitem{Smarr:1977fy}
L.~Smarr, ``{Gravitational Radiation from Distant Encounters and from Headon
  Collisions of Black Holes: The Zero Frequency Limit},'' {\em Phys. Rev. D}
  {\bf 15} (1977) 2069--2077.

\bibitem{Thorne:1980ru}
K.~S. Thorne, ``{Multipole Expansions of Gravitational Radiation},'' {\em Rev.
  Mod. Phys.} {\bf 52} (1980) 299--339.

\end{thebibliography}\endgroup

\end{document}